\documentclass[twocolumn]{aastex61}

\received{--}
\revised{--}
\accepted{--}
\submitjournal{ApJ}

\shorttitle{Atomic gas dominated galaxies at $z\sim$ 0.2}
\shortauthors{Cortese et al.}

\begin{document}
\newcommand{\Zsolar}{\mbox{$\,\rm Z_{\sun}$}}
\newcommand{\Msolar}{\mbox{$\,\rm M_{\sun}$}}
\newcommand{\Lsolar}{\mbox{$\,\rm L_{\sun}$}}
\newcommand{\xs}{$\chi^{2}$}
\newcommand{\dxs}{$\Delta\chi^{2}$}
\newcommand{\xsn}{$\chi^{2}_{\nu}$}
\newcommand{\ls}{{\tiny \( \stackrel{<}{\sim}\)}}
\newcommand{\gs}{{\tiny \( \stackrel{>}{\sim}\)}}
\newcommand{\asec}{$^{\prime\prime}$}
\newcommand{\amin}{$^{\prime}$}
\newcommand{\mstar}{\mbox{$M_{*}$}}
\newcommand{\hi}{H{\sc i}}
\newcommand{\hii}{H{\sc ii}}
\newcommand{\htwo}{H$_{2}$}

\newcommand{\kms}{km~s$^{-1}$\ }


\title{ALMA shows that gas reservoirs of star-forming disks over the past 3 billion years are not predominantly molecular}

\correspondingauthor{Luca Cortese}
\email{luca.cortese@uwa.edu.au}

\author[0000-0002-7422-9823]{Luca Cortese}
\affil{International Centre for Radio Astronomy Research, The University of Western Australia, 35 Stirling Hw, 6009 Crawley, WA, Australia}

\author{Barbara Catinella}
\affil{International Centre for Radio Astronomy Research, The University of Western Australia, 35 Stirling Hw, 6009 Crawley, WA, Australia}

\author{Steven Janowiecki}
\affil{International Centre for Radio Astronomy Research, The University of Western Australia, 35 Stirling Hw, 6009 Crawley, WA, Australia}



\begin{abstract}

Cold hydrogen gas is the raw fuel for star formation in galaxies, and its partition into atomic and molecular phases is a key quantity for galaxy evolution. In this Letter, we combine Atacama Large Millimeter/submillimeter Array and Arecibo single-dish observations to estimate the molecular-to-atomic hydrogen mass ratio for massive star-forming galaxies at $z\sim$ 0.2 extracted from the HIGHz survey, i.e., some of the most massive gas-rich systems currently known. We show that the balance between atomic and molecular hydrogen in these galaxies is similar to that of local main-sequence disks, implying 
that atomic hydrogen has been dominating the cold gas mass budget of star-forming galaxies for at least the past three billion years.
In addition, despite harboring gas reservoirs that are more typical of objects at the cosmic noon, HIGHz galaxies host regular rotating disks with low gas velocity dispersions suggesting that high total gas fractions do not necessarily drive high turbulence in the interstellar medium.

\end{abstract}




\section{Introduction} \label{sec:intro}


It is now well established that, in the local Universe, the vast majority of the cold gas reservoir in galaxies is in the form of atomic hydrogen. On average, only $\sim$30\% of cold gas is found in the molecular phase that directly feeds star formation, though the scatter is large (e.g., \citealp{gass1,coldgass1,hrsco2}). All gas phases rotate in a thin disk with a low degree of turbulence, i.e., velocity dispersions of $\sim$10-30 km s$^{-1}$, depending on the observed phase \citep{epinat2010,ianja2015,mogotsi2016}.

In recent years, cosmological simulations have started providing predictions for the evolution of the molecular-to-atomic hydrogen ratio as a function of redshift. The general expectation is that at earlier cosmic epochs \htwo\ becomes gradually more dominant in galaxies \citep{obreschkow09,lagos11,popping14} and that the gas turbulence increases \citep{genzel11}. This may be related to the fact that high-redshift galaxies are more compact (e.g., \citealp{trujillo2006}) and thus gas surface densities are higher, favoring the condensation of atoms into molecules. With such high densities, the gas would need to be turbulent to avoid gravitational collapse and full consumption by star formation in just a few million years.

Observationally, these predictions have been supported only indirectly by the discovery of star-forming disks at $z\ga$1. These galaxies have stellar masses 
greater than $\sim$10$^{10}$ M$_{\odot}$, and equally large reservoirs of \htwo\ showing a high degree of turbulence (i.e., gas velocity dispersions $\sim$50-60 km s$^{-1}$; \citealp{sins09,swinbank2011,genzel11,genzel13,phibbs2013}). Since the molecular and stellar masses are comparable for these systems, the expectation is that, if \hi\ is present, it must contribute very little (e.g., less than 50\%) to the total cold gas reservoir \citep{genzel15,wuyts2016,burkert16}. Unfortunately, current radio telescopes are unable to detect \hi\ at the cosmic noon, making it impossible to directly prove these assumptions.

Intriguingly, only recently it has become possible to detect \hi\ emission from galaxies at redshift $z\sim$ 0.2-0.4, enabling us to gain some insights into the balance between atoms and molecules beyond our local neighborhood. The data collected so far as part of the CO Observations with the LMT of the Blind Ultra-Deep \hi\ Environment Survey (COOL BUDHIES, \citealp{coolbudhies}) and COSMOS \hi\ Large Extragalactic Survey (CHILES, \citealp{chiles}) show \htwo-dominated gas reservoirs, unlike star-forming galaxies today, lending support to a rapid evolution of the molecular-to-atomic gas ratio as a function of redshift. However, these samples only comprise a handful of galaxies (seven objects in total) and are unlikely to be representative of the star-forming population.

In order to determine whether \hi\ becomes so rapidly negligible beyond the very local Universe, we used the Atacama Large Millimeter Array (ALMA) to measure the \htwo\ masses of galaxies in the HIGHz sample \citep{catinella08,highz}. HIGHz is the highest-redshift sample to date of isolated galaxies with available measurements of \hi\ masses. HIGHz galaxies are massive ($M_{*}>$2$\times$10$^{10}$ \Msolar), disk-dominated systems at $z\sim$0.2 with \hi-to-stellar mass ratios up to 190\% and star formation rates (SFR) between 3 and 35 \Msolar~yr$^{-1}$ (i.e., slightly higher than those typical of main-sequence galaxies at that redshift; \citealp{highz}). As discussed in \S~\ref{results}, the different selection criteria adopted by HIGHz and COOL BUDHIES are such that, once combined, these two samples start providing us with a fairer view of the range covered by the molecular-to-atomic gas ratio of star-forming galaxies at $z\sim$ 0.2.

All of the distance-dependent quantities in this work are computed assuming $\Omega_{M}$=0.3, $\Omega_{\Lambda}$=0.7 and $H_{0}$= 70 \kms Mpc$^{-1}$

\begin{figure*}
\epsscale{1.18}
\plotone{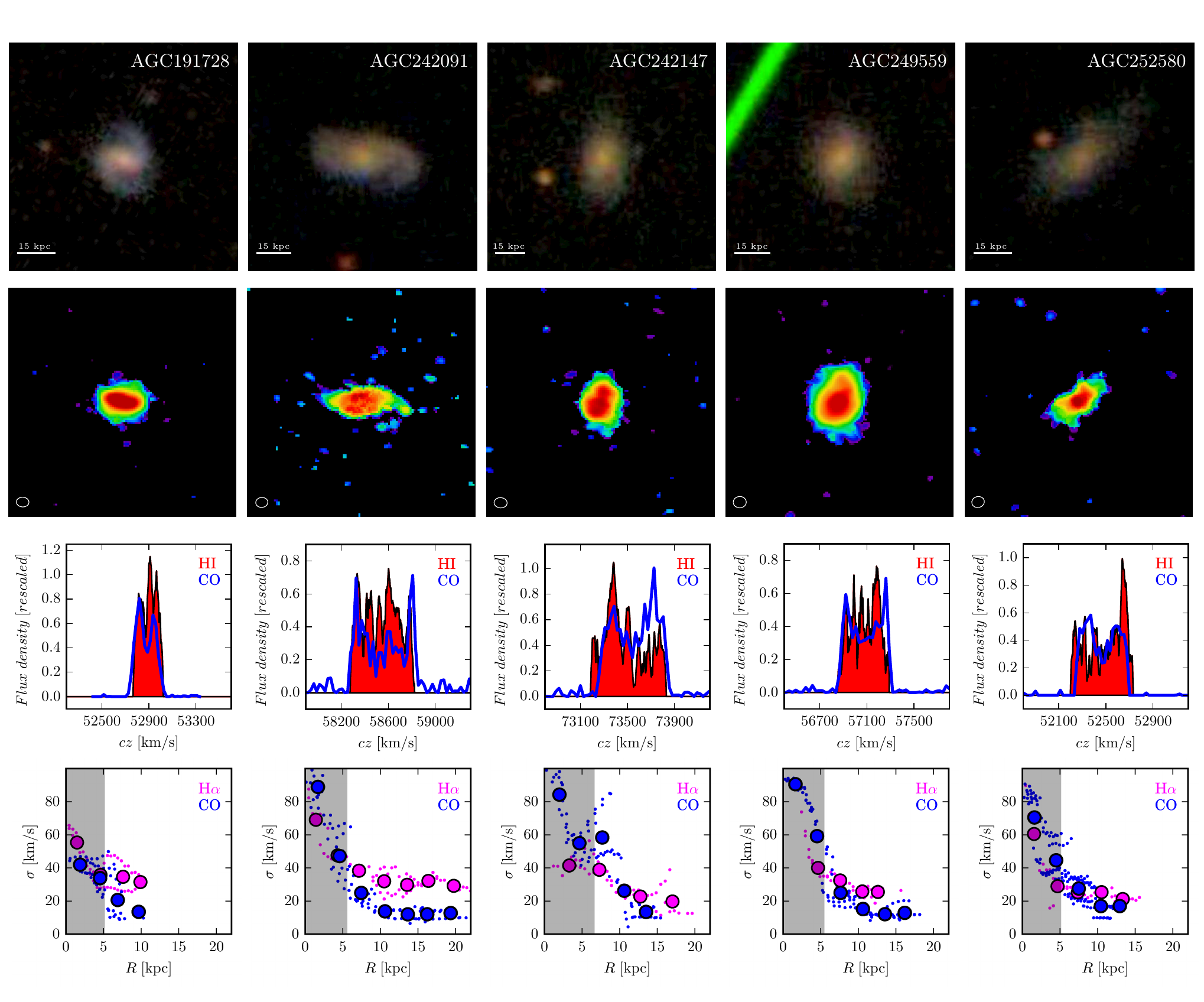}
\caption{Multiwavelength properties of HIGHz galaxies. {\it Top rows}: Optical Sloan Digital Sky Survey color images and ALMA $^{12}$CO(1-0) moment 0 maps. Both sets of images are 30\arcsec wide. The size of the synthesized ALMA beam is shown by the ellipse on the bottom left corner of the $^{12}$CO(1-0) maps.
{\it Third row:} comparison between the integrated Arecibo \hi\ (red) and ALMA $^{12}$CO(1-0) (blue) spectra. Flux densities have been scaled to facilitate the comparison.
{\it Bottom row:} H$\alpha$ (magenta) and $^{12}$CO(1-0) (blue) velocity dispersion as a function of radius. Small points indicate individual measurements, whereas big circles show the averages estimated within 3 kpc wide bins. Averages are presented only when at least four measurements are available. The shaded region corresponds to one FWHM of the ALMA synthesized beam, where measurements are heavily affected by beam smearing. \label{fc}}
\end{figure*}

\section{The Data} \label{sec:data}
\subsection{ALMA $^{12}$CO(1-0) observations} 
We used ALMA to observe five HIGHz galaxies selected to span the range of sizes, \hi\ gas fractions and SFRs 
covered by the entire sample. We imposed an additional declination cut (Declination$<$6\degr) to maximize the observability from the Chajnantor Plateau. HIGHz galaxies were observed as part of program 2015.1.00405 (P.I. Catinella) between 2016 March 5 and 19, during the ALMA observing Cycle 3. The total time on source ranged between 150 and 270 minutes depending on the redshift of the source. We used a similar spectral set-up for all our targets, with four 1875 MHz wide spectral windows, one centered at the expected frequency of the $^{12}$CO(1-0) emission (ranging between 91.3 and 96.8 GHz) and the other three used to obtain continuum observations (not discussed in this Letter). Observatory calibration was used, with either  Callisto, Ganymede, Titan, J0854+2006, J1337$-$1257 or J1550+0527 being observed as flux density calibrator, providing an uncertainty below 10\%. Data were recorded at a spectral resolution of $\sim$3.0 km s$^{-1}$. The data were reduced using the standard ALMA pipeline within the Common Astronomy Software Applications (CASA) package \citep{casa}. Our final data products consist of cleaned flux-calibrated cubes (corrected for variations in the primary beam response) with $\sim$20 \kms channel width ( 
as a trade-off between signal-to-noise and spectral resolution), a synthesized beam of $\sim$1.7 arcsec and a typical rms noise per channel of $\sim$0.2-0.3 mJy beam$^{-1}$. The array configuration used is sensitive to extended emission up to $\sim$20\arcsec. This is larger than the optical size of our targets, suggesting that we are not missing a significant fraction of the total flux.

\begin{deluxetable*}{cccccccc}
\tablecaption{The cold gas content of HIGHz galaxies \label{tab}}
\tablecolumns{8}
\tablewidth{0pt}
\tablehead{
\colhead{AGC\tablenotemark{a}} &
\colhead{SDSS ID} &
\colhead{z$_{SDSS}$} & 
\colhead{$\log$(M$_{*}$)} & 
\colhead{$S_{CO}$} & 
\colhead{$\log$(M$_{H_{2}}$)} &
\colhead{$\log$(M$_{H_{2}}$/M$_{HI}$)} &
\colhead{$\log$(SFR)} \\
\colhead{} &
\colhead{} &
\colhead{} & 
\colhead{M$_{\odot}$} & 
\colhead{Jy km s$^{-1}$} & 
\colhead{M$_{\odot}$} &
\colhead{}&
\colhead{M$_{\odot}$ yr$^{-1}$} 
}
\startdata
191728       &     J091957.00+013851.6     & 0.1763    &   10.89    & 2.91    & 10.15   & $-$0.20  & 1.17 \\
242091       &     J140522.72+052814.6     & 0.1954    &   11.03    & 1.55    & 9.97    & $-$0.61  & 1.24  \\
242147       &     J142735.69+033434.2     & 0.2455    &   11.26    & 2.32    & 10.35   & $-$0.46  & 1.43 \\
249559       &     J144518.88+025012.3     & 0.1906    &   11.17    & 3.86    & 10.34   & $-$0.06  & 1.19 \\
252580       &     J151337.28+041921.1     & 0.1754    &   10.78    & 0.74    & 9.55    & $-$1.04  & 0.79 \\
\enddata
\tablenotetext{a}{Arecibo General Catalog, maintained by M.~P. Haynes and R. Giovanelli at Cornell University}
\end{deluxetable*}

Zeroth-, first- and second-order moment maps have been obtained from the primary-beam corrected cubes as follows. Firstly, cubes are smoothed using a Gaussian filter of width equal to the synthesized beam. Secondly, for each velocity slice, pixels with intensity below three times the typical rms of the smoothed cube are masked. By creating a mask from the smoothed cube we maximize our ability to capture low-level emission that a mask on the original cube would likely miss. Thirdly, moment maps are obtained from the original cube by including only those pixels that are not masked in the smoothed version. The velocity channels included in each moment map have been chosen via visual inspection of each cube slice. 
Alternative ways to extract integrated flux densities include creating moment 0 maps after masking only pixels below 2$\sigma$ rms estimated in the unsmoothed cube, or creating moment 0 maps without masking. All of these methods provide flux densities within 10\% of the value obtained with our preferred technique described above. 

Fig.~\ref{fc} illustrates the main properties of HIGHz galaxies. The top two rows present a comparison between the optical and $^{12}$CO(1-0) images of HIGHz galaxies. The CO emission is spatially resolved and distributed along the disk. The comparison between the \hi\ and CO integrated spectra (third row) highlights the excellent agreement in the velocity range covered by the molecular and atomic phases, implying that both components trace the same gravitational potential. This is far from trivial, given the different spatial resolution of the two datasets, and  provides further confirmation that the \hi\ signal is not significantly contaminated by neighboring galaxies within the Arecibo beam ($\sim$800 kpc diameter at $z$=0.2).


Integrated $^{12}$CO(1-0) luminosities have been estimated from observed flux densities ($S_{CO}$ in Jy km s$^{-1}$) as in \cite{solomon97}:
\begin{equation} 
L_{CO}=3.25 \times 10^{7} S_{CO} \nu^{-2} D_{L}^{2} (1+z)^{-3}
\end{equation}
where $\nu$ is the observed frequency of the $^{12}$CO(1-0) line in GHz, $D_{L}$ is the luminosity distance in Mpc and $z$ is the redshift of the source.    
\htwo\ masses are then obtained by assuming a Galactic CO-to-H$_{2}$ conversion factor of 3.2 \Msolar (K \kms pc$^{2}$)$^{-1}$ (i.e., without including the contribution of Helium). Given that all our galaxies are massive and thus likely metal-rich systems, we can safely assume that a Milky Way conversion factor is a good assumption for our sample. We assume the same CO-to-H$_{2}$ conversion factor for all samples studied in this Letter. This is a conservative assumption, as a lower conversion factor (more typical of extreme outliers of the main sequence of star-forming galaxies, e.g., \citealp{accurso17}) would reinforce our conclusions by further reducing the molecular-to-atomic hydrogen ratio. 
The integrated properties of HIGHz galaxies are presented in Table \ref{tab}.

\subsection{Keck H$\alpha$ spectroscopy}
Optical long-slit spectroscopy of the five HIGHz galaxies presented here has been obtained with the Echellette Spectrograph and Imager (ESI; \citealp{keckESI}) on the 10m Keck II telescope at Maunakea. Observations were carried out on 2016 March 30 and 31, as part of program 2016A-07-Z023E (PI Cortese) in clear sky conditions with $\sim$1 arcsec seeing. We used ESI in echelle mode with slit 0.75\arcsec\ wide and 20\arcsec\ long, and a spectral resolution of R$\sim$5400. We integrated for 4$\times$1200s for AGC242147 and 5$\times$1200s for all the other targets, using position angles aligned to the optical major axis.

\begin{figure*}
\epsscale{1.}
\plotone{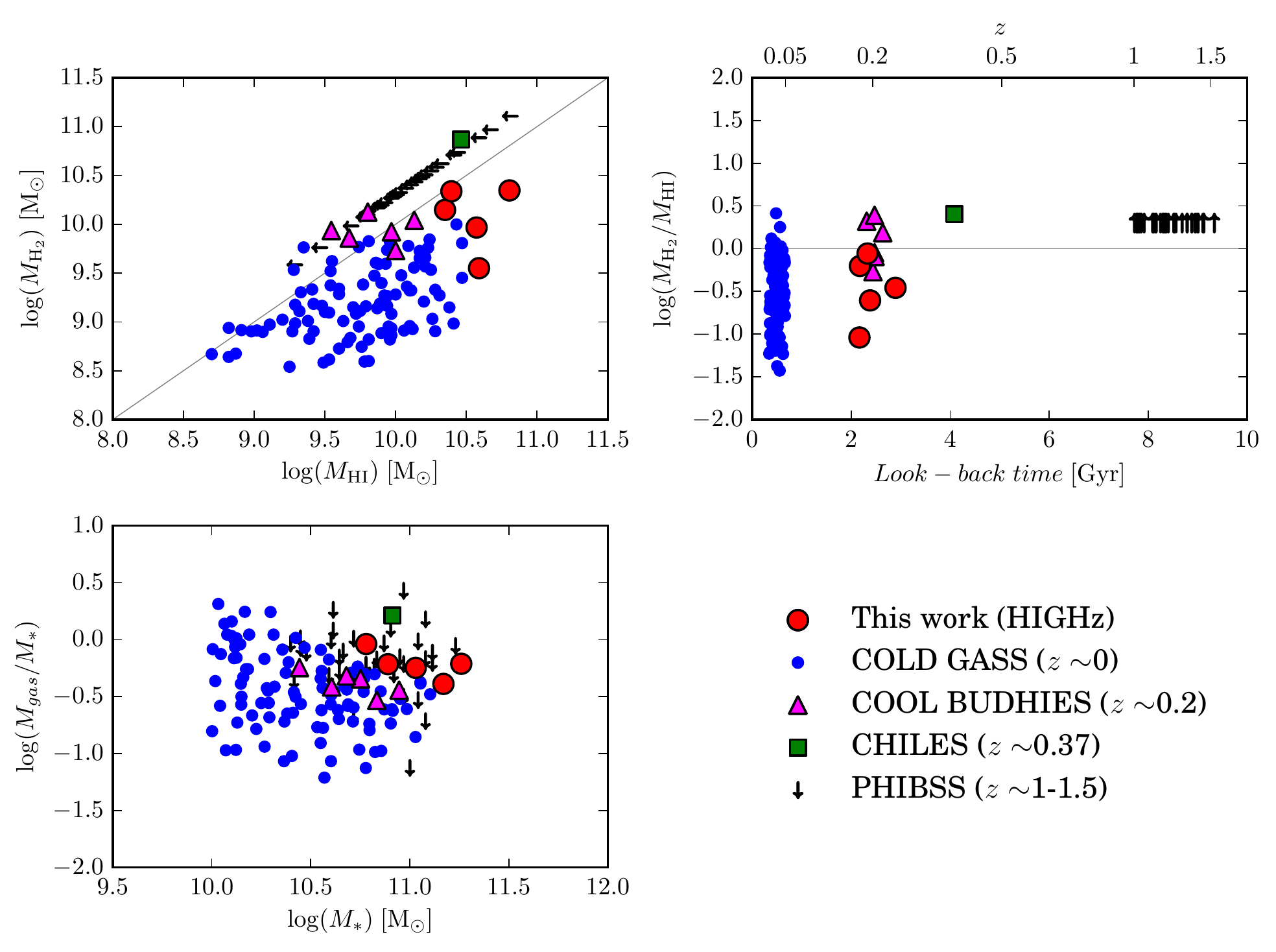}
\caption{Cold gas reservoir of galaxies between $z\sim$ 0 and $\sim$ 1.5. The \htwo\ mass as a function of \hi\ mass (top left), the evolution of the molecular-to-atomic mass ratio with redshift (top right) and the total cold gas fraction as a function of stellar mass (bottom right) are shown. Filled symbols indicate samples with measurements of both atomic and molecular hydrogen. The gray lines in the top panels show the boundary between the \hi- (bottom) and \htwo-dominated (top) regimes. Above $z\sim$ 0.4, detecting \hi\ in emission is unfeasible with current instruments; thus, PHIBSS galaxies are plotted as lower limits assuming $M_{H_{2}}/M_{HI}\geq$2 \label{figratio}.}
\end{figure*}

The spectra were reduced with the standard processing tools in IRAF\footnote{IRAF is distributed by the National Optical Astronomy Observatory, which is operated by the Association of Universities for Research in Astronomy (AURA) under a cooperative agreement with the National Science Foundation.}.
Observations of a bright star were used to trace, straighten, and extract the spectral orders, and calibration lamp spectra were used to generate a wavelength solution. Sky lines were subtracted with a running median in windows away from the galaxy emission. The final velocity resolution ($\sigma$) for each target is $\sim$20 km s$^{-1}$. At this resolution, the nebular H$\alpha$ and [NII] emission lines are cleanly separated. To derive gas velocity dispersions, for each spatial pixel (0.15\arcsec) we extracted a spectrum of the H$\alpha$ line and fit it with a Gaussian profile, rejecting fits with uncertainties in $\sigma$ greater than 15 km s$^{-1}$. For all galaxies we obtained good fits out to at least 3\arcsec\ in radius ($\sim$10 kpc), and in some cases beyond 6\arcsec . While weaker, we also measured other hydrogen and forbidden line profiles as a verification of our H$\alpha$ fits. H$\alpha$ velocity dispersions were then computed by correcting the observed $\sigma$ for both instrumental resolution and redshift broadening.  

\section{Results}\label{results}
The main result of this work is summarized in Fig.~\ref{figratio}, where we compare the \htwo\ and \hi\ content of HIGHz galaxies with those observed in galaxies at different redshifts. 
Panels show the relation between atomic and molecular hydrogen mass (top left), the variation 
of the molecular-to-atomic hydrogen ratio as a function of redshift (top right), and the total (molecular plus atomic, including a factor of 1.3 to account for the contribution of helium) gas-to-stellar mass ratio as a function of stellar mass (bottom-left). Local star-forming galaxies are from the COLD GASS survey (\citealp{gass1,coldgass1}, blue points), whereas samples at redshifts $z\sim$ 0.2 and $\sim$ 0.37 are from the COOL BUDHIES (magenta triangles) and CHILES (green square) surveys, respectively. Black arrows show star-forming galaxies at $z\sim$1.5 from the PHIBSS survey \citep{phibbs2013}. Note that, for consistency with other samples, here we only show COLD GASS galaxies detected in both \hi\ and CO, and lying within 0.5 dex from the main sequence as estimated by \cite{whitaker2012}.
Stellar masses come from either spectral energy distribution fitting (HIGHz, COLD GASS, COOL BUDHIES, PHIBSS) or near-infrared luminosities (CHILES) and, whenever necessary, have been rescaled to a \cite{chabrierIMF} initial mass function.  It is important to note that, while for all the other samples \hi\ flux densities have been directly measured, detecting 21 cm emission at $z\ga$ 0.4 is currently unfeasible. Thus, following previous works (e.g.,\citealp{genzel15,wuyts2016}), we plot PHIBSS galaxies as lower limits assuming $M_{H_{2}}/M_{HI}\geq$2 (but see \S~\ref{concl}). 

All HIGHz galaxies observed by ALMA are \hi-dominated, despite harboring $\sim$10$^{10}$\Msolar\ of \htwo. By combining HIGHz with the COOL BUDHIES sample we show that no significant change in the typical range of molecular-to-atomic hydrogen ratio took place in the past 3 Gyr, and that the gas reservoirs of star-forming galaxies are on average still \hi-dominated. Indeed, the range of values spanned by HIGHz and COOL BUDHIES combined nicely matches that of COLD GASS galaxies. In other words, the addition of HIGHz galaxies clearly reconciles the difference in the molecular-to-atomic hydrogen ratios previously observed between local and higher-redshift objects.

\begin{figure}
\epsscale{1.}
\plotone{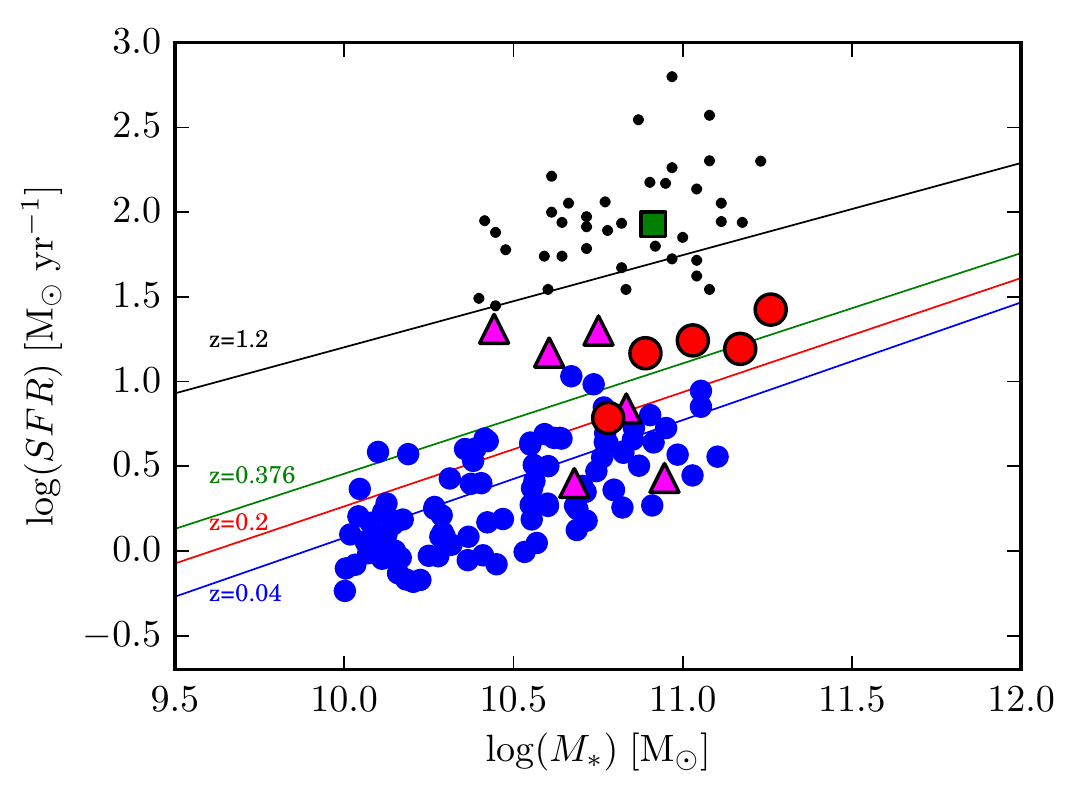}
\caption{The $SFR-M_{*}$ relation for samples at different $z$. Symbols are as in Fig.~\ref{figratio}. Solid lines show the main sequence of star-forming galaxies at each redshift of interest as estimated by \cite{whitaker2012}.\label{sfr}}
\end{figure}

Despite their different $M_{H_{2}}/M_{HI}$ ratio, HIGHz and COOL BUDHIES galaxies have similar SFRs, as shown in Fig.~\ref{sfr}, where 
we present the $SFR-M_{*}$ relations for all samples discussed in this Letter.  
SFR estimates are obtained from a variety of methods including H$\alpha$ luminosities from Sloan Digital Sky Survey spectra (HIGHz), 24$\mu$m luminosities (CHILES), total far-infrared luminosities (COOL BUDHIES), or a combination of ultraviolet and infrared luminosities (PHIBSS and COLD GASS). Solid lines show the location of the main sequence of star forming galaxies at each redshift of interest, as estimated by \cite{whitaker2012}. Due to the lack of homogeneous data, it is impossible to use the same method for all datasets. Nevertheless, these values are only used to compare the position of each sample with respect to its main sequence and show that, on average, HIGHz, COOL BUDHIES, and PHIBSS lie on the upper envelope of their respective sequences. The only clear exception is the CHILES galaxy COSMOS J100054.83+023126.2, which has a SFR almost a factor of 10 higher than what was expected for its stellar mass and redshift.

The different $M_{H_{2}}/M_{HI}$ ratio of HIGHz and COLD BUDHIES is likely due to their different selection criteria. COOL BUDHIES targeted cluster galaxies mainly pre-selected according to their 24$\mu$m emission. Both the environmental and far-infrared selections favor galaxies with a higher than usual molecular-to-atomic ratio, explaining the higher \htwo\ fractions compared to local galaxies \citep{hrsco3,hrsgdratio}. Conversely, HIGHz objects were selected to be isolated systems, avoiding objects with signs of disturbances or dominated by obscured star formation \citep{highz}. These antithetical selection criteria, in addition to the fact that both samples lie close to the main sequence at $z\sim$ 0.2, strongly suggest that, when combined, HIGHz and COOL BUDHIES may show the full range of molecular-to-atomic hydrogen ratio of typical galaxies at $z\sim$ 0.2, which does not appear to be different from that of local galaxies. Unfortunately, with just 11 galaxies and very different selection criteria, it is impossible to quantify any evolution in the {\it average} molecular-to-atomic hydrogen ratio.

Similarly, the high $M_{H_{2}}/M_{HI}$ ratio observed at $z\sim$ 0.37 by CHILES is likely a consequence of the highly obscured SFR of this object. Indeed, even in the local Universe, the most obscured galaxies outliers of the main sequence have more molecular than atomic hydrogen \citep{sanders96}.

HIGHz galaxies have total cold gas masses and gas fractions very similar to those inferred for star-forming turbulent disks at $z\sim$ 1, as shown in the bottom-left panel of Fig.\ref{figratio}. Even if we focus only on the \htwo\ component, there is a significant overlap in gas mass reservoirs between PHIBSS and HIGHz galaxies. This demonstrates that, up to the redshifts where we have been able to measure both \hi\ and \htwo, {\it atomic hydrogen still dominates the cold gas budget even in highly star-forming galaxies}. It also indicates that molecular reservoirs of the order of $\sim$10$^{10}$\Msolar\ do not imply the absence of equally massive \hi\ disks.

Such high total gas fractions have been invoked to explain the large velocity dispersion observed in the interstellar medium (ISM) of $z\sim$ 1 galaxies \citep{genzel11,glazebrook13}. In the bottom row of Fig.~\ref{fc} we combine CO (blue) and H$\alpha$ (magenta) velocity dispersion measurements to test this scenario with HIGHz. We plot single measurements and averages with small and large circles, respectively, as a function of projected distance from the galaxy center. CO velocity dispersions are extracted from moment-two maps using only pixels overlapping with the position of the Keck long-slit. The shaded region in each panel shows the size of the ALMA synthesized beam and highlights where our measurements are heavily affected by beam smearing. 


Our data show that high total gas fractions do not automatically lead to high turbulence in the ISM. The velocity dispersion of the CO component of HIGHz systems is $\sim$10-20 km s$^{-1}$, as observed in local spiral galaxies \citep{mogotsi2016}. Similarly, the H$\alpha$-emitting gas has dispersions $\sim$20-40 km s$^{-1}$, again consistent with that observed in main-sequence star forming galaxies at $z\sim$ 0 \citep{epinat2010}. Indeed, a significant difference between the cold and warm phases of the ISM is in line with previous works \citep{andersen2006}. We note that the velocity 
dispersions obtained for both CO- and H$\alpha$-emitting gas are close to our instrumental resolution. 
Thus, although in most of our objects we reach the flat part of the rotation curve, we cannot exclude that our measurements overestimate the real value. However, this would only reinforce our conclusion that active star formation and high gas richness do not always imply high degree of turbulence.

\section{Conclusion}\label{concl}
In this Letter we combined Arecibo \hi\ single-dish observations with new ALMA $^{12}$CO(1-0) maps of HIGHz galaxies to show that the balance between atomic and molecular hydrogen in massive, gas-rich, star-forming galaxies at $z\sim$ 0.2 is roughly the same as observed in our local neighborhood. Our results demonstrate that previous evidence for a rapid increase of the \htwo-to-\hi\ mass ratio with redshift was likely due to sample selection favoring molecular-dominated systems. Similarly, the low velocity dispersion of both cold and warm components of the ISM in HIGHz systems strongly suggests that it is the local density, rather than the total amount of cold hydrogen, that regulates the level of turbulence of the ISM.  

Given the significant increase in SFR and decrease in typical galaxy size when moving from $z\sim$ 0.2 to $z\sim$ 1, it is still reasonable to expect that the molecular-to-atomic hydrogen mass will increase with redshift. This is particularly true within the optical disk, where the gas surface density should be high enough to allow the condensation of atoms into molecules. However, our results suggest that, given the large scatter in the molecular-to-atomic ratio, \hi\ disks contributing significantly (or even dominating) the total cold gas reservoir of galaxies may still exist in the outer parts of $z\sim$ 1 galaxies. Thus, some caution must be used when assuming that star-forming disks at $z\ga$0.5 have molecular-dominated gas reservoirs.



\acknowledgments
We thank the referee for constructive comments that improved the quality of this Letter. 
We acknowledge support from the Australian Research Council’s Discovery Program (grants FT120100660 and DP150101734). This Letter makes use of the following ALMA data: ADS/JAO.ALMA \#2015.1.00405.S. ALMA is a partnership of ESO (representing its member states), NSF (USA) and NINS (Japan), together with NRC (Canada), NSC and ASIAA (Taiwan), and KASI (Republic of Korea), in cooperation with the Republic of Chile. The Joint ALMA Observatory is operated by ESO, AUI/NRAO and NAOJ. Some of the data presented herein were obtained at the W.M. Keck Observatory, which is operated as a scientific partnership among the California Institute of Technology, the University of California and the National Aeronautics and Space Administration. The Observatory was made possible by the generous financial support of the W.M. Keck Foundation. The authors wish to recognize and acknowledge the very significant cultural role and reverence that the summit of Mauna Kea has always had within the indigenous Hawaiian community. Australian access to the W. M. Keck Observatory has been made available through Astronomy Australia Limited via a scientific collaboration through the Australian National University. Access was supported through the Australian Government's National Collaborative Research Infrastructure Strategy, via the Department of Education and Training, and an Australian Government astronomy research infrastructure grant, via the Department of Industry, Innovation and Science.
%

\vspace{5mm}
\facilities{Arecibo, ALMA, Keck:II (ESI)}

\end{document}